\documentclass[]{aa}  

\usepackage{natbib}
\bibpunct{(}{)}{;}{a}{}{,} % to follow the A&A style
\usepackage{graphicx}
\usepackage{txfonts}
\begin{document}
   \title{Comparing HARPS and Kepler surveys: }

    \subtitle{The alignment  of multiple-planet systems}

   \author{P. Figueira\inst{1}
             \and 
         M. Marmier\inst{2}
         \and
         G. Bou\'{e}\inst{1}
         \and
         C. Lovis\inst{2}
         \and
         N. C. Santos\inst{1,3}
         \and
         M. Montalto\inst{1}
         \and
         S. Udry\inst{2}
         \and 
         F. Pepe\inst{2}
         \and
         M. Mayor\inst{2}
          }

   \institute{Centro de Astrof\'{i}sica, Universidade do Porto, Rua das Estrelas, 4150-762 Porto, Portugal\\
     \email{pedro.figueira@astro.up.pt}
	\and
	 Observatoire Astronomique de l'Universit\'{e} de Gen\`{e}ve, 51 Ch. des Maillettes, 
    - Sauverny - CH1290, Versoix, Suisse
    	\and
	Departamento de F\'{i}sica e Astronomia, Faculdade de Ci\^{e}ncias, Universidade do Porto, Portugal
	}

   \date{}

  \abstract{The recent results of the HARPS and Kepler surveys provided us with a bounty of extrasolar systems. While the two teams extensively analyzed each of their data-sets, little work has been done comparing the two.}{We study a subset of the planetary population whose characterization is simultaneously within reach of both instruments. We compare the statistical properties of planets in systems with $m\sin{i}>$5-10\,M$_{\oplus}$ and R$>$2\,R$_{\oplus}$, as inferred from the HARPS and Kepler surveys, respectively. If we assume that the underlying population has the same characteristics, the different  detection sensitivity to the orbital inclination relative to the line of sight allows us to probe the planets' mutual inclination.}{We considered the frequency of systems with one, two, and three planets as dictated by HARPS data. We used Kepler's planetary period and host mass and radius distributions (corrected from detection bias) to model planetary systems in a simple, yet physically plausible way. We then varied the mutual inclination between planets in a system according to different prescriptions (completely aligned, Rayleigh distributions, and isotropic) and compared the transit frequencies with one, two, or three planets with those measured by Kepler.}{The results show that the two datasets are compatible, a remarkable result especially because there are no tunable knobs other than the assumed inclination distribution. For $m\sin{i}$ cutoffs of 7-10\,M$_{\oplus}$, which are those expected to correspond to the radius cutoff of 2\,R$_{\oplus}$, we conclude that the results are better described by a Rayleigh distribution with a mode of 1\,$^o$ or smaller. We show that the best-fit scenario only becomes a Rayleigh distribution with a mode of 5\,$^o$ if we assume a quite extreme mass-radius relationship for the planetary population.}{These results have important consequences for our understanding of the role of several proposed formation and evolution mechanisms. They confirm that planets are likely to have been formed in a disk and show that most planetary systems evolve quietly without strong angular momentum exchanges such as those produced by Kozai mechanism or planet scattering.}


\authorrunning{P. Figueira et al.}
\titlerunning{On the alignment of multiple-planet systems}

   \maketitle
%
%________________________________________________________________

\section{Introduction}

We live in very exciting times for extrasolar planet science. Since the first discovery of an extrasolar planet by \cite{1995Natur.378..355M}, and thanks to the high efficiency of detection mechanisms, the planetary detection rate has been increasing rapidly. As of today, more than 16 years past, we count $\sim$700 detected planets and many candidates to confirm. We have now both the data and the tools to explore the statistical properties of the planetary underlying population \citep[e.g.][]{2007ARA&A..45..397U}. 

From the technical perspective, during the last couple of years we have witnessed a spectacular increase in the precision (and consequently sensitivity) of the most efficient detection mechanisms. Using the radial velocity (RV) technique, the most efficient to date, the HARPS spectrograph detected 153 planets around G, K, and M stars, amongst which the lightest planets ever found, with a mass of less than 2\,M$_\oplus$ \citep{2009A&A...493..639M, 2011A&A...528A.112L} The only spectrograph with demonstrated sub-m/s precision, HARPS allowed a detailed study of the underlying planetary population, yielding that at least 50\% of the stars of the solar neighborhood have a planet orbiting around them with a period shorter than 100 days \citep{2011arXiv1109.2497M}.

As the Kepler mission became operational and the data were reduced, it became clear that the mission would set a new standard on photometric precision \citep{2010Sci...327..977B}. The candidate extrasolar planets revealed through the transit technique outnumbered those that had been found up to now, from ground or space. Out of the $\sim$1250 candidates found \citep{2011ApJ...736...19B}\footnote{ \cite{2012arXiv1202.5852B},  submitted during the refereeing process of this paper, presented an updated planet count. However, the lack of information on the photometric limits as discussed in first by \cite{2011arXiv1103.2541H} and then by \cite{2011ApJ...742...38Y} for the \cite{2011ApJ...736...19B} release prevents us from repeating the analysis for this new dataset.}, one can highlight the large number of extrasolar planet systems and the first secure confirmation of planets through transit-timing variations \citep{2011Natur.470...53L}.

As information from the two surveys becomes available, a comparison between the results is warranted, and some authors started to tackle this problem \citep[e.g.][]{2011arXiv1108.5842W}. However, this task  is complicated by several factors. The most obvious limitation is that these surveys rely on different techniques that probe different parameter's space. And, naturally, these techniques lead to different detection biases, providing two incomplete snapshots of the underlying population which only overlap partially, giving us a fragmented picture of the mass-radius diagram. 

Particularly interesting is the dependence of both methods on orbital inclination relative to our line-of-sight. Planets transit only when their orbital movement makes them  cover the disk of the star, as seen from our vantage point; this only happens close to our line-of-sight, restraining detections to small angles relative to it. On the other hand, RV techniques probe a much wider range of inclinations: in theory one can detect a system as long as its plane does not coincide with the plane of the sky; in practice the amplitude of the signal depends in a complex way on several orbital parameters, and its detectability is an even more complicated matter.

This has a particularly interesting consequence for planetary systems: while transit technique detects planetary systems that are not only aligned with our line of sight but close to  coplanar, RV detects systems with a much higher inclination and consequently allows the detection of systems with a high mutual inclination between planets. In this work we propose a first comparison between planets from HARPS and Kepler surveys as a way of studying the mutual inclination between planets. We will restrict our analysis to planets with masses and radii within reach of the detection limits of both surveys to allow a simpler and more meaningful comparison.

In Sect.\,2 we present an overview of the HARPS results of interest for our study and in Sect.\,3 we describe our selection and analysis of the Kepler candidates. In Sect.\,4 we describe the methodology that allows us to compare the planets from both surveys and present the results of this comparison. In Sect.\,5 we discuss these results and conclude in Sect.\,6.

%__________________________________________________________________

\section{The HARPS survey results}

\begin{table*}

\caption{The true frequency of single planets and systems with planetary periods between 0.68 and 50 days and $m \sin{i}$ larger than 5, 7, 9, and 10\,M$_\oplus$, as calculated from HARPS data.} \label{tab_HARPS}

\centering
\begin{tabular}{lccccc} \hline\hline
 \ \ {\it m.sin(i)} & Single planets & Syst. w/ 2. plan.  & Syst. w/ 3. plan. & Syst. w/ 4. plan.   \\ \hline \hline

$>$5 \,M$_\oplus$ & 11.15 $\pm$ 3.69 \% &  4.63 $\pm$ 2.27 \% & 0.68 $\pm$ 0.68 \% & n.d.  \\
$>$7 \,M$_\oplus$ & 10.51 $\pm$ 2.60 \% &  2.19 $\pm$ 1.26 \% & n.d. & n.d.  \\
$>$9 \,M$_\oplus$ & 9.82 $\pm$ 2.22 \% &  1.21 $\pm$ 0.85 \% & n.d. & n.d.  \\
$>$10 \,M$_\oplus$ & 8.75 $\pm$ 1.92 \% &  0.57 $\pm$ 0.57 \% & n.d. & n.d.  \\

\hline \hline
\end{tabular}

\medskip

n.d. stands for ``not detected"; see Sect. 2 for more details. 

\end{table*}

The HARPS spectrograph \citep{2003Msngr.114...20M} has set a new benchmark in high-precision RV measurements, delivering several high-visibility results. Recently, \cite{2011arXiv1109.2497M} conducted a detailed statistical analysis of the volume-limited survey of HARPS+CORALIE and calculated the frequency of exoplanets with periods of up to 100 days, by calculating the detection limits for each star and correcting the measured planetary frequency from the detection bias. This analysis was performed on a sample of 822 FGK stars, 376 of which are followed at a level allowing the detection of small-mass planets.

Of interest to us is a subset of the survey, for which comparison with Kepler is possible. We restrain our analysis to exoplanets defined by a mass cutoff whose RV signal makes the detection possible up to the longest period covered by Kepler analysis. This cutoff should be taken with care, since one cannot define {\it a priori} how {\bf large} {\it K}, the semi-major amplitude of the signal, has to be relative to the instrumental precision for the planet to be detected in all cases. Note that this ratio is non-constant and is the reason why to evaluate the presence of planets on datasets one must resort to time-intensive Monte-Carlo analysis as those presented in \cite{2011arXiv1109.2497M}. 

To provide some latitude to our analysis and test the results for different mass-radius relationships, we repeated the analysis of \cite{2011arXiv1109.2497M} for different cutoffs: 5, 7, 9, and 10\,M$_\oplus$ and for the periods between 0.68 and 50 days, the range of periods explored by Kepler. It is important to note that the detection probability of a 5\,M$_\oplus$ planet in a 50 day orbit is already of 10-20\,\% (Fig. 7 of the paper); as a consequence the error bars on the corrected detection rates are quite large, making the results less constraining and less insightful.
These mass-cutoff values will turn out to be a very convenient choice for a different reason, as discussed in the next section. 

The results for the true frequency of single planets and systems with two, three, and four planets are presented in Tab.\,\ref{tab_HARPS}. Note that these results are not corrected for $\sin{i}$ selection effects. This selection effect will bias the detections preferentially toward planets whose system's plane is closer to the line of sight. This might lead to an underestimation of the true frequency of planets, but in a way that is not a function of the number of planets in a system, but rather depends on the detectability of each planet's signal.

\section{The Kepler mission results}

\begin{table*}

\caption{The transit frequency for planets and systems of up to four planets with period between 0.68 and 50 days, as calculated from the Kepler candidates using the methodology from \cite{2011arXiv1103.2541H}.} \label{tab_Kepler}

\centering
\begin{tabular}{lccccc} \hline\hline
 \ \ {\it Radii} &  Single transits & Double transits & Triple transits. & Quadruple transits   \\ \hline \hline
$>$2R$_\oplus$    &  5.23e-1   \%   &    5.97e-2  \% &    9.26e-3 \%  &    5.51e-3  \% \\

\hline \hline
\end{tabular}

\medskip

n.d. stands for ``not detected"; see Sect. 3 for more details.

\end{table*}

The Kepler photometric mission measured high-precision photometric variations of stellar flux on ~160 000 stars (down to $\sim$10 ppm) and allowed (up to now) the detection of 1235 planet transit candidates \citep{2011ApJ...736...19B}. \cite{2011arXiv1103.2541H} extensively analyzed the published datasets and established the properties of the planetary population probed by the mission. Restricting the study to the subset of stars with high planet detectability, the authors quantified the planetary occurrence (as they named the true frequency of planets per star) as a function of planetary radius and period, from 0.68 to 50 days. 

Of particular interest to us are planets with radii large enough for the planetary census to be considered complete and devoid of systematics, the threshold of which was defined as 2\,R$_\oplus$ by the authors.  \cite{2011ApJ...742...38Y} repeated the analysis with a different methodology and, among many interesting results, confirmed this claim of completeness for at least $R>$\,3\,R$_\oplus$. Setting 2\,R$_\oplus$ as the lower limit for data analysis facilitates the comparison with HARPS data, because according to \cite{2007Icar..191..337S} a planet with 10\,M$_\oplus$ has a radius between 1.74 and 2.37\,R$_\oplus$. 

\cite{2011arXiv1103.2541H} presented the planet occurrence as the frequency of planets corrected from both geometrical probability and from the insufficient photometric precision. We repeated the same analysis, considering the same thresholds for detection but now separating planetary candidates into single planets, double transiting systems, triple transiting systems, and quadruple transiting systems.  All planetary and stellar parameters were extracted from the {\it Released Planet Kepler Candidates} list\footnote{http://archive.stsci.edu/kepler/planet\_candidates.html}. We considered the same candidates by imposing the same thresholds on Kepler star magnitude, T$_{eff}$  and log $g$, and setting the same signal-to-noise threshold on transit detection. 
When a system consists of more than one planet we considered the photometric precision to be dictated by the most stringent of the two planets, in line with  \cite{2011arXiv1103.2541H}; to do so we considered the number of stars that still allow us to reach the required precision to detect both planets. 
We corrected our planetary frequencies for the insufficient photometric precision but not for the transit probability. We did this because it is our objective to reproduce through simulations the frequency of planets that transit; these simulations include in a natural way the geometrical probability impact but cannot reproduce the effect of deficient photometric precision.

The frequency of single-transit planets, double transiting systems, triple transiting systems, and quadruple transiting systems is presented in Tab.\,\ref{tab_Kepler}.

%__________________________________________________________________

\section{Comparing the surveys}

\subsection{Methodology}

Comparing the results from the two methods is not straightforward. It is particularly difficult to match a cutoff in radius with one in mass. Here, by using different cutoff values, we considered different mass-radius relationships. This relation is not only unknown, but hardly unique, owing to the different possible composition of exoplanets for the mass range considered \citep[e.g.][]{2007ApJ...665.1413V}.

Starting form the observed HARPS frequencies stated above, we simulated systems whose inclination between the planets and the system's (fixed) plane was controlled by tunable knobs. It is important to note that we considered planet frequencies  as fixed (here used in plural because one must distinguish the frequency of stars with 1, 2, or 3 planets\footnote{In principle, there is no reason not to consider the frequency of stars with more planets around them; in practice these systems were not present in the sample.}). These were dictated by HARPS results and only depend on the mass cutoff (i.e. the mass-radius relationship assumed). The methodology is depicted in the flowchart presented in Fig.\,\ref{flowchart} and can be summarized in the following way:

\begin{figure*}
\centering
\includegraphics[width=12.5cm]{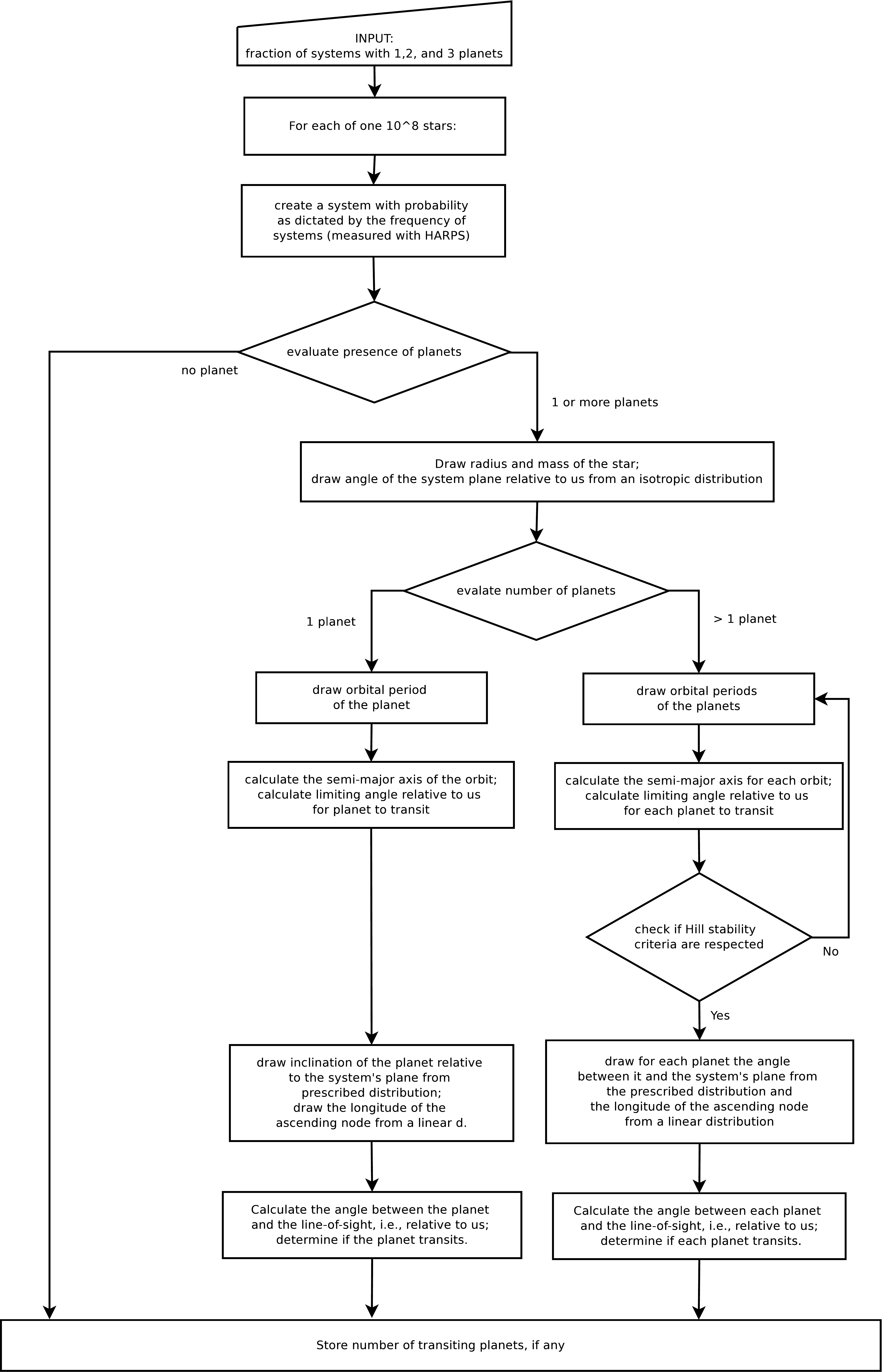}
\caption{Flowchart for the simulations described in Sect. 4.1.}\label{flowchart}
\end{figure*}

\begin{itemize}

\item we created 10$^8$ planetary systems; the number of planets in each system was dictated by the frequencies of stars with one, two, or three planets set by HARPS results.

\item for each system the stellar radius was drawn from Kepler's radius distribution of host stars. Each planet period $P_k$ was drawn from the measured Kepler period distribution of planetary candidates. 

The observed stellar radius distribution and planetary period distribution were corrected both from photometric detectability bias and from geometry bias. The geometry bias is corrected simply by applying the formula of the transit probability \citep[e.g.][]{2011ApJ...742...38Y}

\begin{equation}
	P_{{\rm transit}} = 0.051\left( \frac{10\,{\rm days}}{P}  \right) ^{2/3} \left( \frac{\rho_\odot}{\rho_*} \right) ^{1/3} \, ,
\end{equation}

in which $\rho$ is the mean stellar density and is valid for the cases of $R_p\ll R_*$ and e\,=\,0 \footnote{We considered that planetary orbits are circular, a hypotheses strengthened by the work of \cite{2011ApJS..197....1M}, who showed that Kepler transit durations imply a mean eccentricity $\le$ 0.2.}. The photometric detectability bias, as corrected by \cite{2011arXiv1103.2541H}, reduces the accuracy of our measurements because it is presented for binned data in the ($\log{P}$, $\log{R_p}$) space. To correct for this last effect we used instead the power-law formulas (4), (5), and (6) of \cite{2011ApJ...742...38Y}. The interested reader is referred to this work for the details on the advantages of the approach. We corrected for these two effects and only then binned the data into 20 equally sized bins and assigned a probability proportional to the (corrected) frequency of planets inside the bin; these distributions are presented in Fig.\,\ref{distributions}.

\begin{figure}

\includegraphics[width=9.5cm]{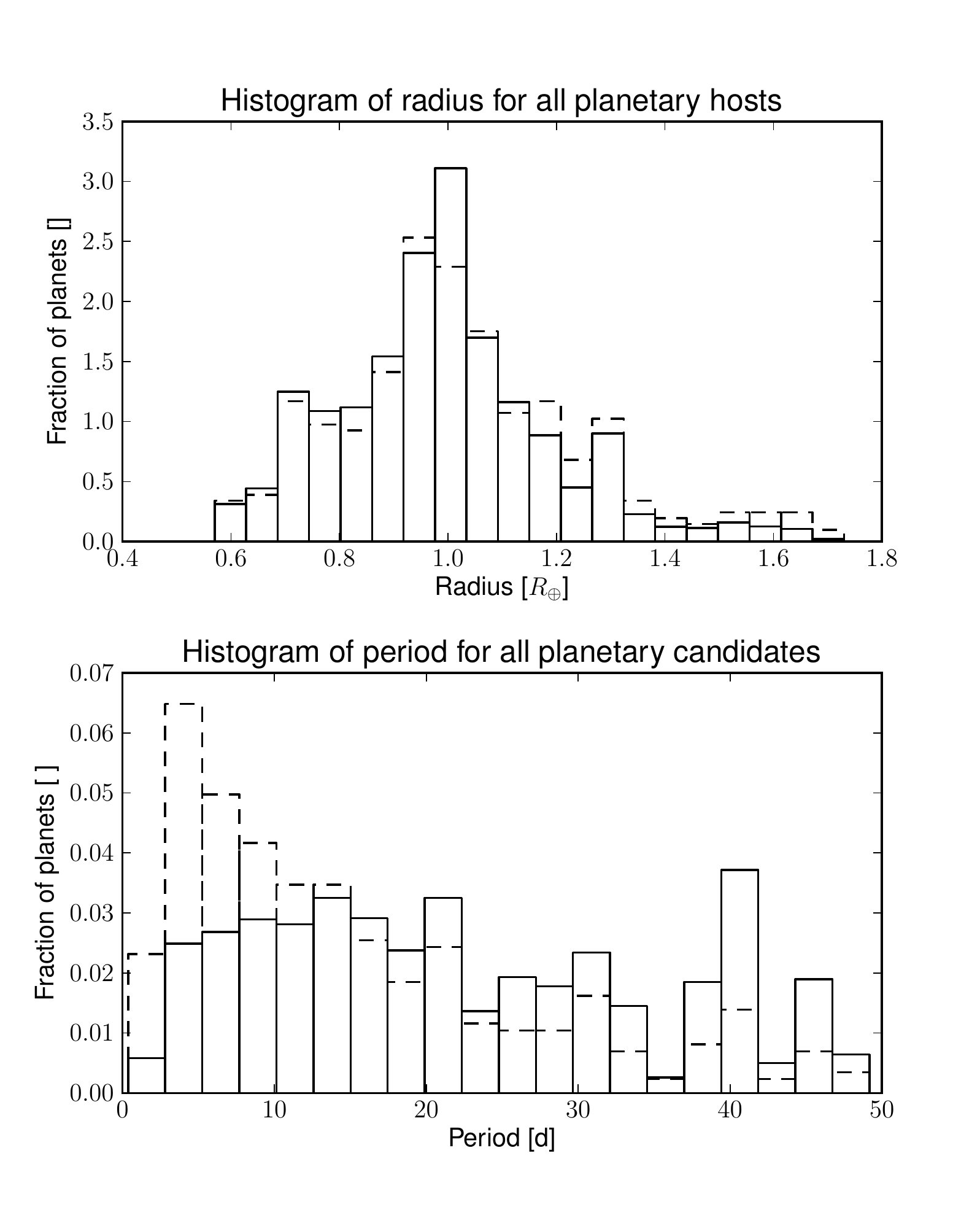}

\caption{Distribution of radii for all Kepler planetary host ({\it top}) and period for all Kepler planetary candidates considered for the 2\,$R_{\oplus}$ radii cutoff({\it bottom}); dashed histograms depict the distributions prior to the detection bias correction and solid line after. }\label{distributions}

\end{figure}

\item For each drawn stellar radius, we drew a stellar mass with a flat probability between the extreme masses detected for the radii included in the bin. This allows us to translate orbital periods  $P_k$ into orbital semi-major axis $a_k$ using the generalized Kepler third law:

\begin{equation}
a_i = \sqrt[3]{\frac{M_*}{M_\odot} \frac{P}{1\,yr}^2} [A.U.] \,.
\end{equation}

\item If more than one planet is drawn for the system in question, the orbits have to respect the Hill stability criterion, as defined by Eq. (24) of \cite{1993Icar..106..247G}, 

\begin{equation}
\frac{a_2-a_1}{a_1} > 2.40 \, (\mu_1+ \mu_2)^{1/3}\label{Gladman} \, , 
\end{equation}

in which $\mu_1$=$m_1$/$m_*$, $\mu_2$=$m_2$/$m_*$. This equation is valid for circular orbits and when $m_* \gg m_1, m_2$. The planetary periods are redrawn until all pairs of planets respect this property. We chose for $m_1$ and $m_2$ the $m\sin{i}$ cutoff used for the considered HARPS frequency determination. 

\item The orientation of the  system's plane relative to the line-of-sight plane, {\it I}, is drawn from a linear distribution in {\it sin} (note that angles closer to the line of sight have a higher probability of being detected). 

\item For each planet the angle $i_k$ with respect to the reference plane of the system is drawn from the assumed distribution (see next subsection), and the longitude of the ascending node $\Omega_k$ is drawn from a linear distribution between 0 and 360$^o$. 

\item Finally, each angle $\theta_k$ relative to the line of sight, as seen from our vantage point, is given by

\begin{equation}
\theta_k = {\rm asin}\,|\,\cos{I}\cos{\Omega}\sin{i} +\sin{I}\cos{i}\,| \, ,
\end{equation}

and the planet is considered to transit if $\theta_k<\theta_{lim} = {\rm asin}(R_*/a_k)$. The different planes and angles mentioned are depicted in Fig.\,\ref{figure_coord}.

\begin{figure}

\includegraphics[width=8cm]{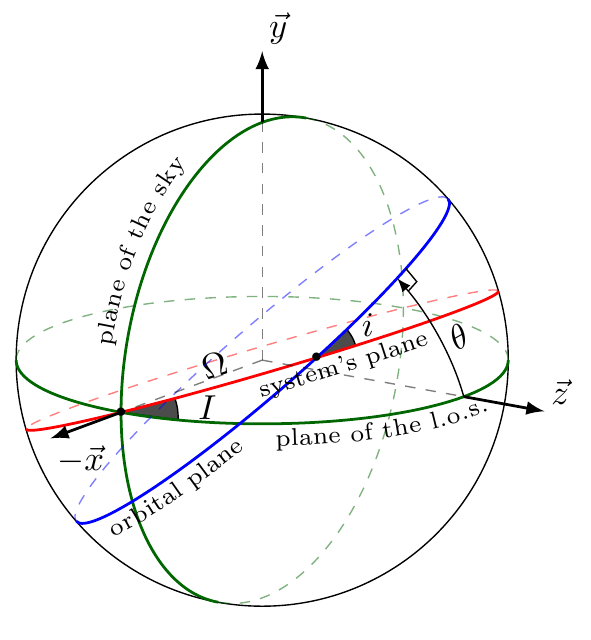}

\caption{Angles used in this paper. The observer is in the $\vec{z}$ direction.}\label{figure_coord}

\end{figure}

\end{itemize}

It is important to note that we followed a methodology similar to that of \cite{2011ApJS..197....8L}, in which one starts from different primordial populations and tries to recover the frequency of observed Kepler transit candidates. However, in our case, and on top of several different minor aspects, we had a previously determined frequency of planets in systems for single planets, double-planet systems, and triple-planet systems. To estimate the error bars on our transit frequencies we used the 1-$\sigma$ uncertainties in HARPS frequencies to calculate extreme-case scenarios for the transit frequencies. 

\subsection{Mutual inclination distributions and results}

For the distribution of the angles of each planet relative to the plane of the system {\it i} we considered $1)$ planets aligned with their systems plane; $2)$ different Rayleigh distributions {\it R}($\sigma$) relative to the system's plane, where $\sigma$, the mode of the distribution, is the only governing parameter. For the latter we considered $\sigma\,\in$\,[1, 5, 10, 20, 30, 45]$^o$; and $3)$ an isotropic distribution (i.e. a linear distribution in $\cos{i}$). The results are presented in Tab.\,\ref{transit_frequencies}.

\begin{figure*}

\centering

\includegraphics[width=15cm]{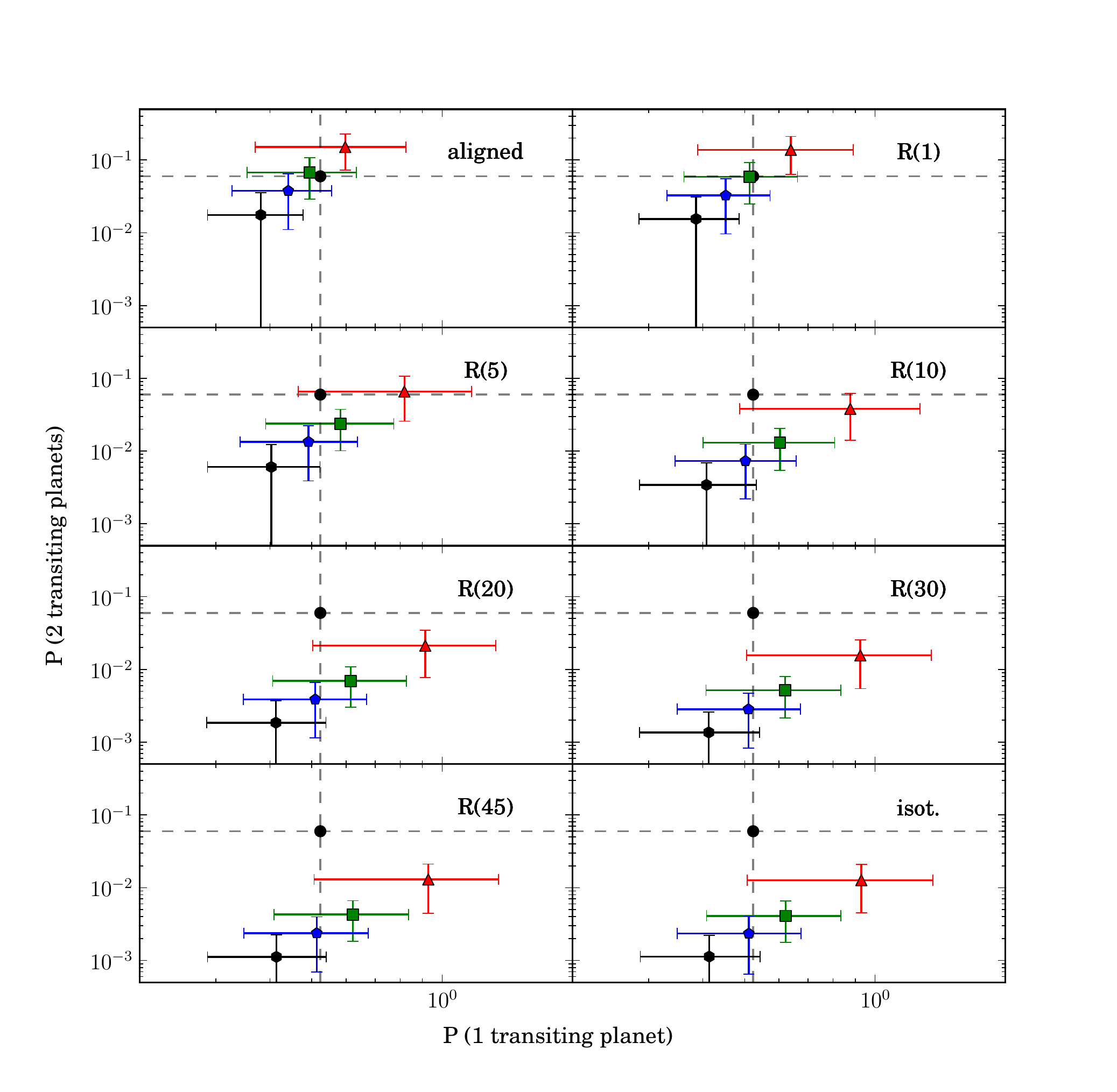}

\caption{Result of the simulations with the methodology described in 4.1 for the different {\it i} distributions considered.}\label{simulations}

\medskip

The figure depicts the match between the frequency of single and double transiting systems obtained from our simulations and Kepler results. In each plot the black circle represents Kepler results and the triangle, square, pentagon, and hexagon (red, green, blue, and black in electronic version only) represent the results for the 5, 7, 9, and 10\,$M_{\oplus}$ cutoffs in $m\sin{i}$; the error bars result from the application of the 1-$\sigma$ uncertainties. The different subpanels represent the application of different {\it i} distributions, from left to right and top to bottom: aligned, {\it R}(1$^o$), {\it R}(5$^o$), {\it R}(10$^o$),  {\it R}(20$^o$),  {\it R}(30$^o$), {\it R}(45$^o$), and isotropic, respectively.

\end{figure*}

The obtained frequencies for single and double planetary transits are plotted and compared to those recovered from Kepler data in Fig\,\ref{simulations}. In this figure we plot in a two-dimensional space the transit frequencies (and associated uncertainties) for single-transiting planets and double-transiting planets, as calculated from our simulations, for the different assumed inclinations (different panels) and different $m\sin{i}$ cutoff (different markers and colors). In the same plot we present the values measured from Kepler data, so that both can be directly compared. 

Note that for most cutoffs considered HARPS did not detect triple or quadruple systems for the given mass and period range. This was due to 1) the low number of stars surveyed when compared with Kepler's, for which the detection of a system with frequency of $10^{-3}$ is quite low, and 2) to the short period range considered in the analysis, of 50 days. As a matter of fact, many of the systems detected by HARPS extend to more than 50 days, up to 100 days, and were naturally not included in the analysis.

\begin{table*}

\caption{Frequency of simulated transiting systems with 1, 2, and 3 planets, for the different planetary frequencies of Tab.\ref{tab_Kepler} }\label{transit_frequencies}

\centering

\begin{tabular}{cccccc} \hline\hline

 \ \  $m\sin{i}$ cutoff & distribution  & Single planets [\%] & Syst. w/ 2. plan.  [\%]  & Syst. w/ 3. plan.  [\%] \\ \hline \hline
      	
5 \,$M_{\oplus}$ &  aligned & 5.97e-01$_{-2.27e-01}^{+2.27e-01}$ & 1.51e-01$_{-7.79e-02}^{+7.80e-02}$ & 1.93e-02$_{-1.88e-02}^{+1.88e-02}$ \\[4pt]
 5 \,$M_{\oplus}$ &  {\it R}(1.0$^o$) & 6.39e-01$_{-2.50e-01}^{+2.52e-01}$ & 1.38e-01$_{-7.47e-02}^{+7.30e-02}$ & 1.42e-02$_{-1.37e-02}^{+1.41e-02}$ \\[4pt]
 5 \,$M_{\oplus}$ &  {\it R}(5.0$^o$) & 8.18e-01$_{-3.53e-01}^{+3.52e-01}$ & 6.60e-02$_{-4.02e-02}^{+4.10e-02}$ & 2.31e-03$_{-1.81e-03}^{+2.40e-03}$ \\[4pt]
 5 \,$M_{\oplus}$ &  {\it R}(10.0$^o$) & 8.77e-01$_{-3.90e-01}^{+3.93e-01}$ & 3.81e-02$_{-2.40e-02}^{+2.37e-02}$ & 7.17e-04$_{-2.17e-04}^{+6.53e-04}$ \\[4pt]
 5 \,$M_{\oplus}$ &  {\it R}(20.0$^o$) & 9.14e-01$_{-4.12e-01}^{+4.16e-01}$ & 2.12e-02$_{-1.35e-02}^{+1.33e-02}$ & 2.05e-04$_{-2.05e-04*}^{+2.04e-04}$ \\[4pt]
 5 \,$M_{\oplus}$ &  {\it R}(30.0$^o$) & 9.25e-01$_{-4.20e-01}^{+4.25e-01}$ & 1.56e-02$_{-1.01e-02}^{+9.80e-03}$ & 1.07e-04$_{-1.07e-04*}^{+1.11e-04}$ \\[4pt]
 5 \,$M_{\oplus}$ &  {\it R}(45.0$^o$) & 9.29e-01$_{-4.23e-01}^{+4.21e-01}$ & 1.30e-02$_{-8.54e-03}^{+8.10e-03}$ & 6.30e-05$_{-6.30e-05*}^{+5.70e-05}$ \\[4pt]
 5 \,$M_{\oplus}$ &  isotropic & 9.30e-01$_{-4.23e-01}^{+4.30e-01}$ & 1.27e-02$_{-8.18e-03}^{+8.30e-03}$ & 6.90e-05$_{-6.90e-05*}^{+4.40e-05}$ \\[2pt]

\hline\hline
 
 7 \,$M_{\oplus}$ &  aligned & 4.94e-01$_{-1.40e-01}^{+1.39e-01}$ & 6.75e-02$_{-3.86e-02}^{+4.05e-02}$ & n. d.\\[4pt]
 7 \,$M_{\oplus}$ &  {\it R}(1.0$^o$) & 5.13e-01$_{-1.51e-01}^{+1.49e-01}$ & 5.88e-02$_{-3.38e-02}^{+3.39e-02}$ & n. d.\\[4pt]
 7 \,$M_{\oplus}$ &  {\it R}(5.0$^o$) & 5.82e-01$_{-1.91e-01}^{+1.91e-01}$ & 2.38e-02$_{-1.37e-02}^{+1.38e-02}$ & n. d.\\[4pt]
 7 \,$M_{\oplus}$ &  {\it R}(10.0$^o$) & 6.03e-01$_{-2.02e-01}^{+2.04e-01}$ & 1.31e-02$_{-7.64e-03}^{+7.40e-03}$ & n. d.\\[4pt]
 7 \,$M_{\oplus}$ &  {\it R}(20.0$^o$) & 6.15e-01$_{-2.09e-01}^{+2.11e-01}$ & 6.94e-03$_{-3.92e-03}^{+3.96e-03}$ & n. d.\\[4pt]
 7 \,$M_{\oplus}$ &  {\it R}(30.0$^o$) & 6.20e-01$_{-2.13e-01}^{+2.13e-01}$ & 5.15e-03$_{-3.00e-03}^{+2.83e-03}$ & n. d.\\[4pt]
 7 \,$M_{\oplus}$ &  {\it R}(45.0$^o$) & 6.22e-01$_{-2.13e-01}^{+2.15e-01}$ & 4.28e-03$_{-2.45e-03}^{+2.35e-03}$ & n. d.\\[4pt]
 7 \,$M_{\oplus}$ &  isotropic & 6.22e-01$_{-2.14e-01}^{+2.12e-01}$ & 4.09e-03$_{-2.32e-03}^{+2.50e-03}$ & n. d.\\[2pt]

\hline\hline

 9 \,$M_{\oplus}$ &  aligned & 4.41e-01$_{-1.14e-01}^{+1.14e-01}$ & 3.77e-02$_{-2.66e-02}^{+2.66e-02}$ & n. d.\\[4pt]
 9 \,$M_{\oplus}$ &  {\it R}(1.0$^o$) & 4.52e-01$_{-1.21e-01}^{+1.20e-01}$ & 3.26e-02$_{-2.29e-02}^{+2.28e-02}$ & n. d.\\[4pt]
 9 \,$M_{\oplus}$ &  {\it R}(5.0$^o$) & 4.91e-01$_{-1.50e-01}^{+1.46e-01}$ & 1.34e-02$_{-9.50e-03}^{+8.90e-03}$ & n. d.\\[4pt]
 9 \,$M_{\oplus}$ &  {\it R}(10.0$^o$) & 5.02e-01$_{-1.57e-01}^{+1.56e-01}$ & 7.31e-03$_{-5.12e-03}^{+5.09e-03}$ & n. d.\\[4pt]
 9 \,$M_{\oplus}$ &  {\it R}(20.0$^o$) & 5.09e-01$_{-1.62e-01}^{+1.60e-01}$ & 3.87e-03$_{-2.72e-03}^{+2.79e-03}$ & n. d.\\[4pt]
 9 \,$M_{\oplus}$ &  {\it R}(30.0$^o$) & 5.10e-01$_{-1.61e-01}^{+1.63e-01}$ & 2.83e-03$_{-2.00e-03}^{+1.90e-03}$ & n. d.\\[4pt]
 9 \,$M_{\oplus}$ &  {\it R}(45.0$^o$) & 5.13e-01$_{-1.65e-01}^{+1.62e-01}$ & 2.37e-03$_{-1.68e-03}^{+1.61e-03}$ & n. d.\\[4pt]
 9 \,$M_{\oplus}$ &  isotropic & 5.11e-01$_{-1.62e-01}^{+1.63e-01}$ & 2.34e-03$_{-1.69e-03}^{+1.70e-03}$ & n. d.\\[2pt] 

\hline\hline
 
 10 \,$M_{\oplus}$ &  aligned & 3.81e-01$_{-9.40e-02}^{+9.60e-02}$ & 1.76e-02$_{-1.71e-02}^{+1.79e-02}$ & n. d.\\[4pt]
 10 \,$M_{\oplus}$ &  {\it R}(1.0$^o$) & 3.86e-01$_{-1.01e-01}^{+9.90e-02}$ & 1.55e-02$_{-1.50e-02}^{+1.55e-02}$ & n. d.\\[4pt]
 10 \,$M_{\oplus}$ &  {\it R}(5.0$^o$) & 4.03e-01$_{-1.16e-01}^{+1.19e-01}$ & 6.04e-03$_{-5.54e-03}^{+6.26e-03}$ & n. d.\\[4pt]
 10 \,$M_{\oplus}$ &  {\it R}(10.0$^o$) & 4.08e-01$_{-1.22e-01}^{+1.24e-01}$ & 3.43e-03$_{-2.93e-03}^{+3.50e-03}$ & n. d.\\[4pt]
 10 \,$M_{\oplus}$ &  {\it R}(20.0$^o$) & 4.13e-01$_{-1.27e-01}^{+1.26e-01}$ & 1.85e-03$_{-1.35e-03}^{+1.87e-03}$ & n. d.\\[4pt]
 10 \,$M_{\oplus}$ &  {\it R}(30.0$^o$) & 4.13e-01$_{-1.27e-01}^{+1.28e-01}$ & 1.36e-03$_{-8.60e-04}^{+1.25e-03}$ & n. d.\\[4pt]
 10 \,$M_{\oplus}$ &  {\it R}(45.0$^o$) & 4.14e-01$_{-1.27e-01}^{+1.26e-01}$ & 1.12e-03$_{-6.20e-04}^{+1.14e-03}$ & n. d.\\[4pt]
 10 \,$M_{\oplus}$ &  isotropic & 4.14e-01$_{-1.27e-01}^{+1.29e-01}$ & 1.13e-03$_{-6.30e-04}^{+1.07e-03}$ & n. d.\\[2pt]

 \hline\hline

\end{tabular}

\medskip 

The error bars were drawn from the lower and upper limit of HARPS frequencies uncertainties. n.d. stands for ``not detected"; see Sect. 4 for more details. Note that for very low frequency values (marked with a *), lower than 5$e$-04\%, the number of events is so low that the error bars are ill defined through Monte-Carlo analysis and cannot be considered meaningful.

\end{table*}

\section{Discussion}

\subsection{The likeliness of different {\it i} distributions}

The first point to note is that the simulations based on HARPS frequencies agree remarkably well with the Kepler measurements. This is worth noting, especially because our simulation procedure has no tunable knobs; no effort was made to match the calculated with the observed data. The only free parameter is the inclination distribution, as described.

\begin{figure}

\includegraphics[width=9cm]{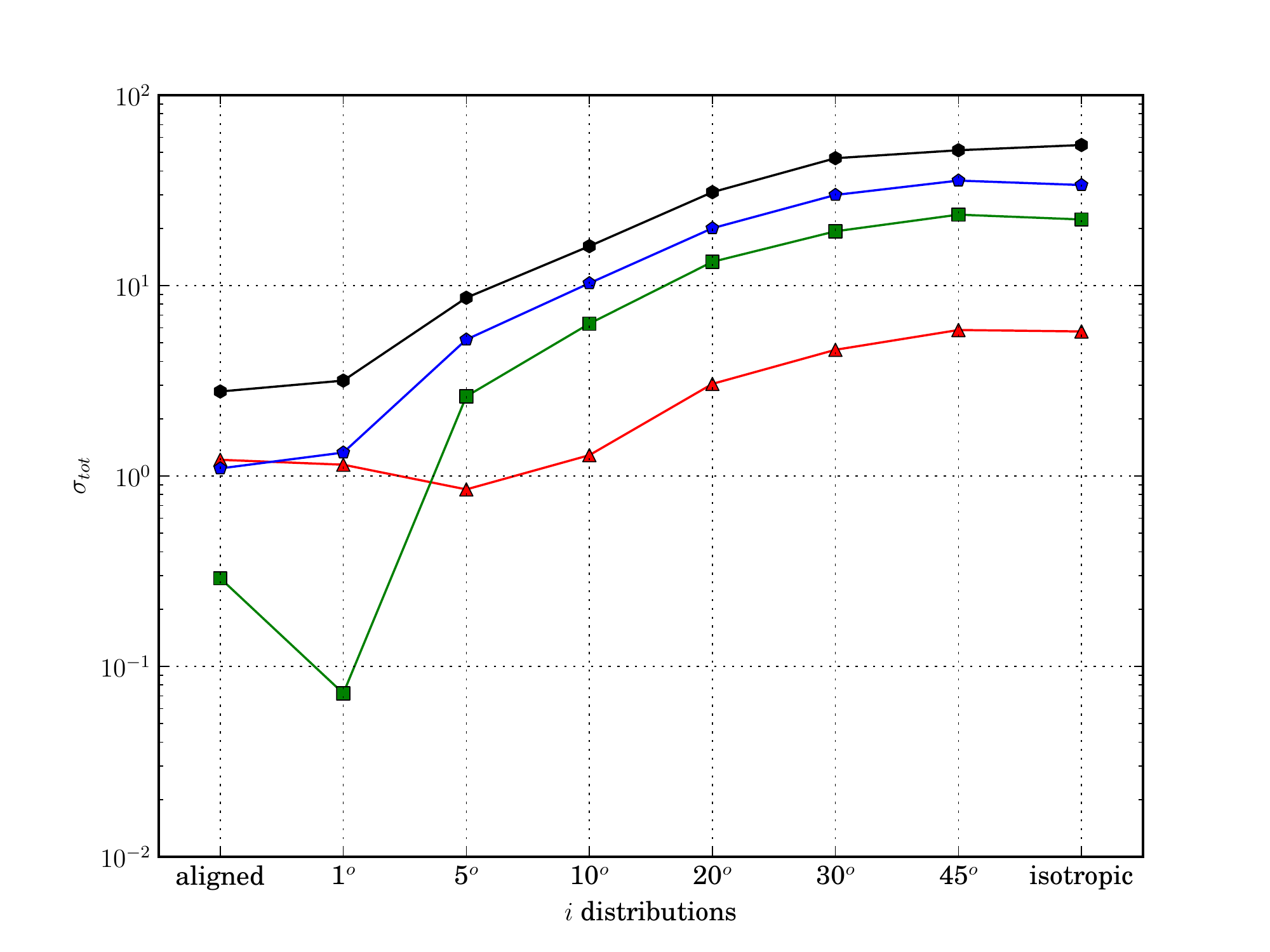}

\caption{Distance between Kepler point and the result from each simulation, measured in $\sigma$'s; the triangle, square, pentagon, and hexagon (red, green, blue, and black in electronic version only) represent the results for the 5, 7, 9, and 10\,$M_{\oplus}$ cutoffs in $m\sin{i}$, respectively.  }\label{i_distributions}

\end{figure}

From Fig.\,\ref{simulations} one can conclude that while all considered distributions reproduce the fraction of stars with one transiting planet inside or close to 1-$\sigma$ error bars, most underestimate the frequency of stars with two transiting systems. It is interesting to quantify the difference between the observed quantities and those reproduced through simulations. To do so we calculated for each case the absolute deviation between the observed and calculated values, taking as units the 1-$\sigma$ uncertainties (our proxy for the transit frequencies' $\sigma$). The results are presented in Tab.\,\ref{comparison}, and in Fig.\,\ref{i_distributions} we plot the quadratic sum of the two absolute deviations $\sigma_{tot}=\sqrt{\sigma_{f_1}^2 + \sigma_{f_2}^2} $, as a function of the considered $i$ distribution for the different $m\sin{i}$ cutoffs.

\begin{table*}

\caption{The absolute deviation between the Kepler transit frequencies for single and double transit systems for the 2\,R$_\oplus$ cutoff and those obtained through simulation in this paper (presented in Tab.\,\ref{transit_frequencies}) for the different assumed inclination distributions.} \label{comparison}

\centering
\begin{tabular}{cccccccccc} \hline\hline
 \ \  &  \multicolumn{2}{c}{O-C(5\,$M_{\oplus}$) }  & \multicolumn{2}{c}{O-C(7\,$M_{\oplus}$)} & \multicolumn{2}{c}{O-C(9\,$M_{\oplus}$)} & \multicolumn{2}{c}{O-C(10\,$M_{\oplus}$)}  \\ 
 \ \ {\it i} dist. &  $\sigma_{f_1}$  & $\sigma_{f_2}$ &  $\sigma_{f_1}$  & $\sigma_{f_2}$ &  $\sigma_{f_1}$  & $\sigma_{f_2}$ &  $\sigma_{f_1}$  & $\sigma_{f_2}$  \\ 
 \hline \hline
aligned &  0.33 &  1.17 &  0.21 &  0.20 &  0.72 &  0.83 & 1.48 & 2.35  \\
 {\it R}(1.0$^o$) &  0.46 &  1.05 &  0.07 &  0.03 &  0.59 &  1.19 & 1.38 & 2.85  \\
{\it R}(5.0$^o$) &  0.84 &  0.16 &  0.31 &  2.60 &  0.22 &  5.20 & 1.01 & 8.57  \\
{\it R}(10.0$^o$) &  0.91 &  0.91 &  0.40 &  6.30 &  0.13 &  10.29 & 0.93 & 16.08  \\
{\it R}(20.0$^o$) &  0.95 &  2.89 &  0.44 &  13.32 &  0.09 &  20.01 & 0.87 & 30.94  \\
{\it R}(30.0$^o$) &  0.96 &  4.50 &  0.46 &  19.28 &  0.08 &  29.93 & 0.86 & 46.67  \\
{\it R}(45.0$^o$) &  0.96 &  5.77 &  0.46 &  23.58 &  0.06 &  35.61 & 0.87 & 51.39  \\
isotropic &  0.96 &  5.66 &  0.46 &  22.24 &  0.07 &  33.74 & 0.84 & 54.74  \\

\hline \hline
\end{tabular}

\end{table*}

An inspection of the Tab.\,\ref{comparison} and Fig.\,\ref{i_distributions} shows that, in general, the more misaligned the planets (the larger the mode of the {\it i} distribution considered), the stronger the deviation relative to the measured Kepler values. For cutoffs of 9 and 10\,$M_{\oplus}$ this is exactly the case, but for cutoffs of 5 and 7\,$M_{\oplus}$ the most probable distributions are not the aligned one but Rayleigh with modes of 1 and 5\,$^o$, respectively. This brings back the question of the impact of the $m\sin{i}$ cutoff on our results. We saw in Sect. 3 that a 2\,$R_{\oplus}$ radius corresponds to a mass cutoff of around 10\,$M_{\oplus}$, of which precise value depends on the composition of the planets. We also know that by using $m\sin{i}$ as a proxy for the real mass, we are underestimating the mass by a factor of 0.73, the average value of the $\sin{i}$ for an isotropic distribution. Accordingly, for the $m\sin{i}$ cutoffs of 5, 7, 9, and 10, we are considering (on average) mass cutoff values of 6.3, 8.9, 11.4, and 12.7\,$M_{\oplus}$, respectively. At face value, this means that the results for $m\sin{i}$ of 7 and 9\,$M_{\oplus}$ are those that better match the cutoff in mass; interestingly, the result for the $m\sin{i}$ cutoff of 7\,$M_{\oplus}$ indeed provides the best match, and is twice as close to Kepler's measured values than the second best. However, the question is more involved than that: the span in composition means that a planet with mass lower by a couple of \% than the average mass cutoff has a non-zero probability of having a radius above the cutoff. This sends a clear warning about matching a sharp cutoff in mass with one in radius.

It is therefore important to note that only for an extreme choice of mass cutoff we have the (relatively) high Rayleigh mode of 5\,$^o$ as the best match. As we saw before, one would have to assume a mass cutoff significantly lower than 8.9\,$M_{\oplus}$ and closer to 6.3\,$M_{\oplus}$, which is at odds with the structure model's results. According to \cite{2007Icar..191..337S}, a planet with mass of 6.3\,$M_{\oplus}$ has a radius between 1.66 and 2.09\,$R_{\oplus}$, for a water content of $<$0.1\% and $\sim$50\%, respectively. This shows that planets with such mass are indeed too small to be included, or at least to contribute significantly for the population we recover by imposing as limiting radius 2\,$R_{\oplus}$. The only exception are strongly irradiated planets, as discussed by \cite{2011ApJ...738...59R}, for which, under very strong irradiation (500\,$<$T$_{eff} <$\,1000\,K) and other assumptions, a planet with mass as low as 4\,$M_{\oplus}$ can have a radius large enough to be included in our cutoff. However, once again, it is unlikely that these planets dominate our population.

It is important to note that Eq.\,\ref{Gladman} is only valid for coplanar systems. Multi-planetary systems with relative inclinations higher than zero have more degrees of freedom and are thus more easily unstable. Then, stability should require larger planet separations. Nevertheless, in this study we chose to treat all systems in the same way to avoid any uncontrolled bias. Since planets on wider orbits have lower transit probabilities, our choice slightly increases the transit probability of the outer planets of non-coplanar systems. The same argument can be used to show that if larger masses are used in the calculation of the Hill stability criterion, the spacing between the planets is increased and the transit probability decreased. Unfortunately, owing to the lack of appropriate stability studies for inclined systems, one cannot do better than we do here.
 
We can then conclude that the inclination distribution that better reproduces Kepler single and double-transit frequencies follow a Rayleigh distribution with mode of $\sim$1\,$^o$ or even smaller, but we caution that the limitations of the approach presented here lead more to an order-of-magnitude result than to a clearly defined value. What can be asserted from our analysis is that a Rayleigh distribution with a mode of 5\,$^o$ can only be accommodated if we consider a population characterized by an extreme mass-radius relationship that would lead to very light planets ($\sim$6\,$M_{\oplus}$) with a typical radius larger than 2\,$R_{\oplus}$. 
 
\subsection{The impact of different stellar hosts: spectral types and metallicity}

When comparing the outcome of different surveys, one must take into account that the population of stars surveyed even though similar, has different characteristics. The first point to evaluate is the spectral type of the hosts themselves, since planetary formation is expected to be a function of stellar mass \citep[e.g.][]{2011A&A...526A..63A}, among other parameters. 

\begin{figure}

\includegraphics[width=8cm]{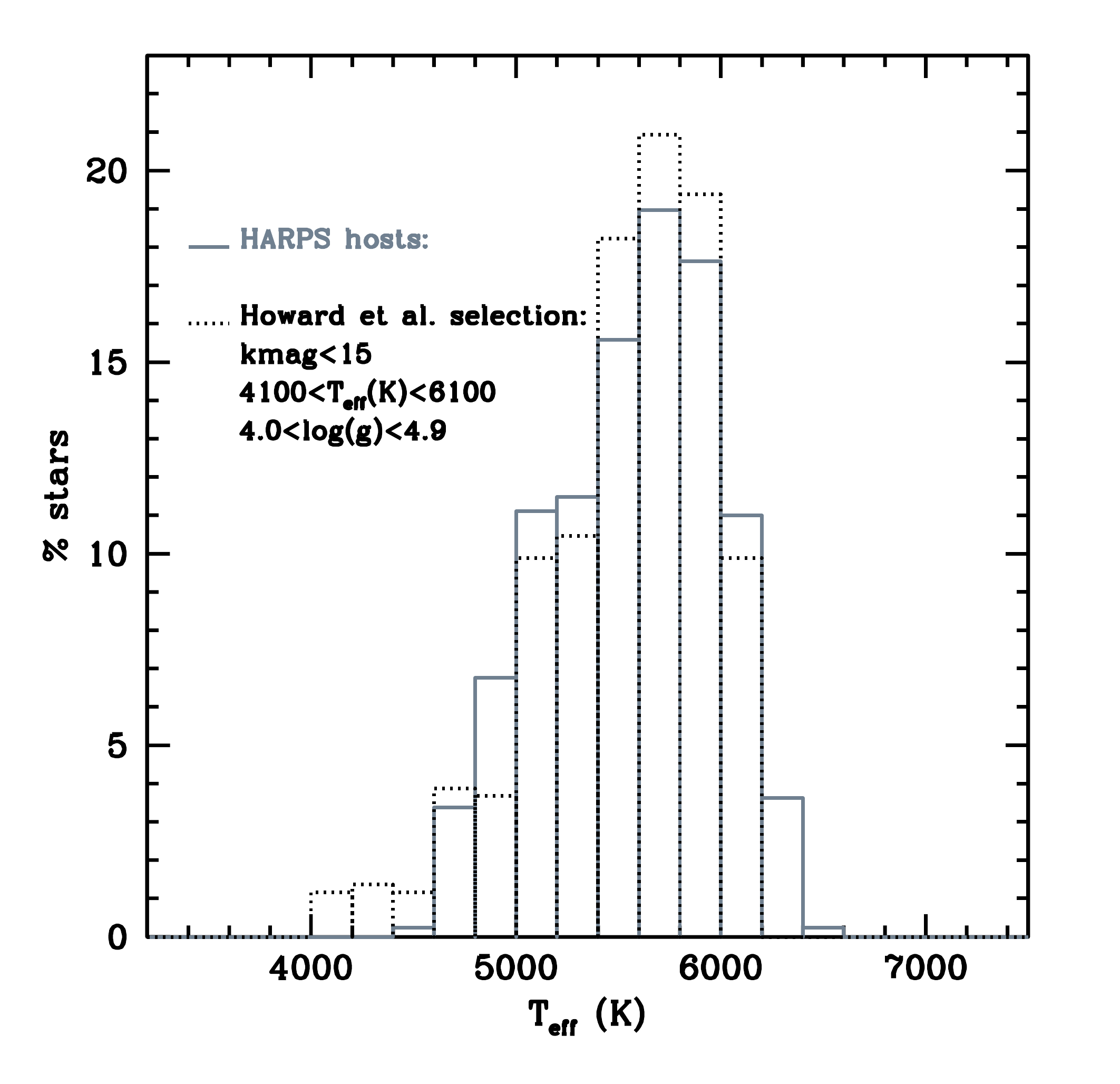}

\caption{Histogram of Kepler stars selected by effective temperature, from 3000 to 7500\,K in bins of 200\,K. In gray and with a solid line we present HARPS host and in black with a dotted line we present stars selected from the Kepler candidate stars according to the \cite{2011arXiv1103.2541H}  selection criteria (which are thus eligible for planetary detection).}\label{plot_hist_host}

\end{figure}

In Fig.\,\ref{plot_hist_host} we overplot the two T$_{eff}$ distributions obtained from HARPS and Kepler hosts considered in this study. We notice that Kepler host temperatures extend up to 400\,K cooler than the coolest HARPS hosts, and conversely HARPS hosts extend to 400\,K hotter than the hottest Kepler hosts. Both distributions have their peak around 5700\,K, but the Kepler distribution is more peaked while HARPS is wider. 

The T$_{eff}$ for HARPS survey was determined as described by the calibration of \cite{2008A&A...487..373S} and its internal accuracy is of 50\,K. The accuracy of the effective temperatures given by the Kepler Input Catalog (KIC) was estimated to be within 200 K for stars in the range between 4500 K and 6500 K \citep{2011AJ....142..112B}. Recently, systematic differences between the griz magnitudes and those of the SDSS (DR8) have been noticed \citep{2011arXiv1110.4456P}. Comparisons of (J-Ks)-based temperatures from the infrared-flux method (IRFM) and SDSS filter indicated a mean shift towards hotter temperatures with respect to the KIC {\bf of} the order of 215 K in between 4000 K and 6500 K. This result seems to be confirmed also by high-precision spectroscopic measurements. \cite{2010ApJ...723.1583M} determined the effective temperature of the sun-like star KIC11026764 by means of several spectroscopic approaches. They determined that this object is typically 100 K and 300 K hotter than the value reported in the KIC, which is in line with the possible offset between our two distributions.

Another possible bias in our comparison may come from the different metallicity distribution of the HARPS and Kepler samples. On the one side, all HARPS stars are in the solar neighborhood (closer than 50 pc), and the vast majority belong to the Galactic thin disk. On the other hand, Kepler targets are all much farther away (no longer solar neighborhood objects). Adding to this, Kepler targets are slightly above the Galactic plane \citep{2011arXiv1103.2541H}, suggesting that they may be more metal-poor (on average) than solar neighborhood stars because of a higher proportion of thick disk objects\footnote{Thick-disk stars are more deficient on average than thin-disk counterparts \citep[e.g.][]{2011A&A...535L..11A}.}. This may have important implications regarding the type of planets found in the two samples. For instance, it has been shown that giant planets are preferentially found around metal-rich stars \citep{2001A&A...373.1019S, 2004A&A...415.1153S,  2005ApJ...622.1102F}, while this same trend is not seen for neptunes and super-Earths \citep{2007ARA&A..45..397U, 2011arXiv1109.2497M, 2011A&A...533A.141S}. The stellar metal content may also affect the stellar radius, since [Fe/H] correlates with stellar radius \citep{2006A&A...453L..21G}. The impact of these effects is difficult to quantify; the two quantities are expected to be correlated. However, using the Besan\c{c}on model, \cite{2011arXiv1103.2541H} derived a $\sim$0.1\,dex effect in [Fe/H] as a function of spectral type, a value that is not very pronounced.

The implications of these differences are far from obvious and their study requires a detailed knowledge of the impact of stellar host properties on planet formation -- and eventually a more refined analysis of Kepler hosts parameters. This is far beyond the reach of this paper, and we will content ourselves with noting that the two populations do not seem to differ in a significant way. 

\subsection{Comparison with other works and consequences for formation and evolution mechanisms} 

\cite{2011ApJS..197....8L} analyzed the architecture of Kepler planetary system candidates (i. e. the observed multiplicity frequencies) and concluded that a single population of planetary systems that matched the higher multiplicities underpredicted the number of single-transiting systems. The authors also provided constraints on the frequency of systems with 1 to 6+ planets and mutual inclination, even though these two parameters are naturally correlated.
However, their results were obtained using a methodology that required normalizing the number of simulated transits to reproduce the total number of transits detected by Kepler, making it vulnerable to the presence of false planet positives (FPP). \cite{2011ApJ...738..170M}  calculated a FPP rate $<$ 10\% for 90\% of all Kepler candidates, the average being closer to 5\% and with tails extending up to 30\%. The presence of FPP also has an impact on our study, since it will modify the planetary period distribution and stellar radii and mass distributions. However, and unlike \cite{2011ApJS..197....8L}, here the distinction between the different models is made essentially through the frequency of double transiting extrasolar planets, for which the FPP is much lower than on average or for single-planet systems \citep{2012arXiv1201.5424L}, which makes our results probably more robust.

\cite{2012AJ....143...94T} described a very general formalism to analyze the multiplicity function in transit and RV surveys based on the approximation of separability, that the probability of distribution of planetary parameters in a system is the product of identical one-planet distributions. The authors applied it both to the Kepler survey and to RV+Kepler data (in which they used the planets listed in {\it exoplanet.org} as of August 2010 for RV data) and reached several interesting conclusions. They found out that a wide range of inclinations were compatible with the Kepler data alone, as long as some of the stars contained many planets per system \citep[curiously different from the conclusions of ][but still in agreement with their upper limit on an rms inclination of 10\,$^o$]{2011ApJS..197....8L}.
Their joint analysis of RV and transit data led to an rms inclination between 0 and 5\,$^o$, a result compatible with ours. Their wider range of inclinations can be explained by their different working hypothesis, different RV data, and especially by a more general and less detailed treatment of detection biases for both surveys.

The very recent work of \cite{2012arXiv1202.5852B} presents a new list containing 1091 new Kepler exoplanet candidates based on the analysis of an extended data-set covering 16 months. However, and as already discussed before, the lack of information on the photometric limits as discussed first by \cite{2011arXiv1103.2541H} and then by \cite{2011ApJ...742...38Y} for  the \cite{2011ApJ...736...19B} release prevents us from repeating our analysis for the new dataset. Still, some educated guesses can be made on its impact on our results. The first point to note is that the highest fractional increases are obtained for small-radius candidates (197\% for candidates smaller than 2\, $R_{\oplus}$ compared to 52\% for candidates larger than it) and located at longer orbital periods (123\% for planets outside of 50 day orbits versus 85\% for candidates inside of 50 day orbits) and thus not for the population discussed here. If we apply to this dataset the same selection criteria we applied to \cite{2011ApJ...736...19B}, the number of stars with one transiting candidate increase from 274 to 384 (a 40\% increase) and the stars with two transiting candidates increase from 30 to 41 (a 37\% increase). Since the match of observations to the models in our study is governed mostly by the frequency of stars with two transiting planets, we expect that a higher frequency of stars with two transiting planets will favor more aligned models. As discussed in \cite{2012arXiv1202.5852B}, these frequencies will increase, owing to the higher number of events and the usage of a more precise pipeline. However, without applying the detectability correction coming from photometric limits we cannot translate the absolute number of events into the required frequencies and draw firm conclusions. We leave this analysis to a future paper.

The recently submitted paper of \cite{2012arXiv1202.6328F} revisits the architecture of Kepler multi-transit systems and discussed several of their most important properties. Among the addressed points is the distribution of inclinations for the candidates, which the authors estimate to be between 1 and 2.3\,$^o$. The authors conclude then that the new release of Kepler data alone allows one to establish that planetary systems are strongly coplanar and declare that studies such as \cite{2012AJ....143...94T} and our own, which draw information from RV studies too, are unnecessary complicated by the RV surveys' different characteristics. However, as demonstrated by \cite{2012AJ....143...94T}, the transit data alone cannot exclude, for instance, the presence of two population of planetary systems: one whose relative inclination is low and other population with different properties, and potentially higher planet relative inclinations and planetary frequencies. In other words, it is important to keep in mind that \cite{2012arXiv1202.6328F} evaluate the inclination of {\it planetary systems with multi-transiting planets}, while the studies that rely on RV and Kepler evaluate the inclination of {\it  planetary systems as whole}. The two values are not derived for the same population of planets and cannot be compared directly.

Our work shows that planetary systems are likely to host planets with a very low inclination relative to the plane of the system. This is already the case for the solar system, which has an average inclination $<$\,2\,$^o$, and favors the standard model for planet formation in a disk. More importantly, it suggests that most planets in systems do not have their orbital elements influenced by violent angular momentum exchanges such as planet-planet scattering \citep[e.g.][]{2008ApJ...678..498N}, Kozai oscillation \citep[e.g.][]{2003ApJ...589..605W} or perturbation by a stellar encounter \citep[e.g.][]{2011MNRAS.411..859M}, mechanisms that are only expected to create short-period single planets and not systems.  As already pointed out by Greg Laughlin\footnote{Oral communication at the ``First Kepler Science Conference", December 2011.}, the distribution of orbital parameters for Kepler multi-transit systems is very similar to that of the same (scaled) parameters for Solar System giant planet satellites. This is well in line with the fact the planets should be formed in a disk, with relative low orbital inclinations, as we show here.

%__________________________________________________________________

\section{Conclusions}

We attempted at a first comparison between the planetary population properties as characterized by the HARPS and Kepler surveys. We simulated the population of planets, with planetary frequencies dictated by HARPS survey results, and considered that these planets followed different inclination distributions relative to the systems' plane. We considered distributions from aligned to Rayleigh distributions with different modes, to finally isotropic (completely independent).

The first remarkable point is the compatibility of the results from the two surveys. This is made even more so by the very little freedom we have on influencing our simulations' outcome. There are no tunable knobs other than the inclination distribution. Concerning these, we showed that the results point to a strong alignment of the systems, with an inclination between the planet and the plane of the system better described by a Rayleigh distribution with mode of $\sim$1\,$^o$ or even lower. This is a feature that depends on the assumed mass-radius relationship, but not in a strong way, an important point since the mass-radius relationship for low-mass planets is ill-defined and hardly unique because of the span of compositions and mass that can generate the same radius. We have shown that only in extreme and consequently highly unlikely cases the best fit for the alignment between planets is expected to follow an $R$(5$^o$).

These results have important consequences for our understanding of the role of several proposed formation and evolution mechanisms. They confirm that planets are likely to have been formed in a disk and show that most planetary systems evolve quietly without strong angular momentum exchanges such as those produced by the Kozai mechanism or planet scattering.

\begin{acknowledgements}
This work was supported by the European Research Council/European Community under the FP7 through Starting Grant agreement
number 239953, as well as by Funda\c{c}\~ao para a Ci\^encia e a Tecnologia (FCT) in the form of grant reference
PTDC/CTE-AST/098528/2008. NCS would further like to thank FCT through program Ci\^encia\,2007 funded by FCT/MCTES (Portugal) and POPH/FSE (EC). MM is supported by grant SFRH/BPD/71230/2010, from FCT (Portugal). PF would like to thank to Manuel Monteiro for his highly skilled and prompt assistence on running the simulations in a cluster. 

\end{acknowledgements}

% for the bibliography, at the end
\bibliographystyle{aa} % style aa.bst
\bibliography{Mybibliog} % your references Yourfile.bib

\end{document}